\documentclass[10pt,journal,letterpaper,twocolumn,twoside]{IEEEtran} 

\usepackage{amsmath,amsfonts,bm,amssymb,psfrag,ifthen,color,subfigure}
\usepackage{mathrsfs}
\usepackage[justification=centering,font=footnotesize]{caption}

\usepackage[IEEEtran]{research17}
\allowdisplaybreaks[4]
\usepackage{bbm}
\newcommand{\C}{\const{C}}

\usepackage[dvips]{epsfig}
\usepackage[dvips]{graphicx}
\usepackage{amsmath,amsfonts,bm,amssymb,psfrag,ifthen,color,subfigure}
\usepackage{algpseudocode}
\usepackage{algorithm}
\usepackage{algorithmicx}
\usepackage{ifthen}
\usepackage{setspace}
\usepackage{cite}
\usepackage{url}
\usepackage{longtable}
\usepackage{stfloats}  
\usepackage{graphics,booktabs,color,subfigure}
\usepackage{float}
\usepackage{doi}
\usepackage{color}

\author{\IEEEauthorblockN{%
          Shuai Ma\IEEEauthorrefmark{1}\IEEEauthorrefmark{2}\IEEEauthorrefmark{3} and Michèle Wigger\IEEEauthorrefmark{1}}\\%
  \IEEEauthorblockA{\IEEEauthorrefmark{1}LTCI, Telecom Paris,
	IP Paris, 91120 Palaiseau, France}\\%
  \IEEEauthorblockA{\IEEEauthorrefmark{2}National Mobile Communications Research Laboratory, Southeast University, Nanjing 210096, China}\\
  \IEEEauthorblockA{\IEEEauthorrefmark{3}School of Information and
	Control Engineering, China University of Mining and Technology, Xuzhou
	221116, China}\\
  \IEEEauthorblockA{mashuai001@cumt.edu.cn, michele.wigger@telecom-paris.fr}\\%
}

\usepackage{listings}

\floatname{algorithm}{Algorithm}

\begin{document}
\newcommand{\mw}[1]{{\color{blue}#1}}

\title{First- and Second-Moment Constrained Gaussian Channels}

\maketitle
\begin{abstract}

This paper studies the channel capacity of intensity-modulation direct-detection (IM/DD) visible light communication (VLC)    systems
  under both     optical and electrical power constraints.
  Specifically,  it derives    the asymptotic capacities in the  high and low signal-to-noise ratio (SNR) regimes under  peak, first-moment,  and second-moment constraints. The
results show that     first- and second-moment constraints are never  simultaneously active in the asymptotic low-SNR regime, and only in few cases in the  asymptotic high-SNR regime. Moreover, the second-moment constraint is more stringent in the asymptotic low-SNR regime than in the high-SNR regime.

\end{abstract}

%

\IEEEpeerreviewmaketitle

\section{Introduction}

The ever-increasing number of wireless devices and high-speed communication requirements  cause a spectrum scarcity of  conventional radio-frequencies (RF).  A promising solution
is  visible light communication (VLC) with its  abundant unlicensed spectrum \cite{Jovicic,Pathak}.
  In particular, when utilizing the simple and practical intensity modulation--direct detection (IM/DD)  technology, transmitters directly modulate information
onto the   real,  non-negative   optical intensity of the VLC signals (in contrast to RF signals which modulate  the complex field) and receivers apply photodetectors to measure incoming optical intensities.
For eye safety  reasons and   hardware limitations,
 both the
  maximum and   average optical  intensities of VLC transmit signals  typically have to be  restricted.
   Since these apply directly  to the intensities, they impose both peak and \emph{first-moment constraints} on the transmit signals.
Additional \emph{second-moment constraints} are imposed by limitations of the electronic circuits that control the transmit signal, such as the boundedness of the linear amplification regime and electric power consumption \cite{Standard,Shuai_hang,Ling,Huang,Che,Ma}.

A close-form expression for the capacity of such IM/DD systems is still unknown, even when some of the first or second-moment constraints are relaxed. However, bounds and asymptotic results in the high- and low signal-to-noise ratio (SNR) regimes are known under certain relaxations.
For example,
 various upper and lower bounds   on the capacity,
 as well as its exact  high- and low-SNR asymptotics, have been derived  under only a first-moment constraint without a second-moment constraint   \cite{Lapidoth_1,Chaaban_1,Wang_1,Chan_1,Jiang_1,Farid_1,Farid_2}.

 In this work, we derive the exact expressions for the asymptotic high- and low-SNR capacities  under peak,   first-moment,  \emph{and}  second-moment constraints. Our results show that in the asymptotic low-SNR regime, only one of the two moment constraints is stringent. Specifically, the second-moment constraint is active  if the peak-constraint $\amp$ times the first-moment-constraint $\alpha_1 \amp$ exceeds this second-moment constraint $\alpha_2 \amp^2$, and otherwise the first-moment constraint is active. This can be seen as a  consequence of the optimality of on-off keying in the asymptotic low-SNR regime.   Our results further show that for most constraint-parameters $(\alpha_1,\alpha_2)$ also in  the  high-SNR regime, only one of the moment-constraints is active. Interestingly, the second-moment constraint is inactive over a larger region of  $(\alpha_1,\alpha_2)$-pairs in the high-SNR regime than in the low-SNR regime, and the first-moment constraint  over a smaller region.   An additional  second-moment constraint is thus more restrictive in the low-SNR regime than in the high-SNR regime. In the asymptotic high-SNR regime,  we further observe a small region of  $(\alpha_1,\alpha_2)$-pairs where both moment-constraints are simultaneously active and limit the asymptotic capacity.

\section{Channel Model}
Consider a typical VLC communication link, where the transmitter is equipped
with a single LED or laser and the receiver with  a single  photodetector.
The photodetector measures the incoming light intensity, which can be  modeled as
\begin{IEEEeqnarray}{c}
  Y =  {x}  + Z,
  \label{eq:channel}
\end{IEEEeqnarray}
where   ${x}$  denotes the  input signal produced by the transmitter's LED or laser, and $Z$ is standard additive white   Gaussian noise independent of ${x}$.
Note that, in contrast to the input ${x}$, the output $Y$ can be negative.

Inputs ${x}$ are subject to both a peak   and an
average optical power (average-intensity) constraints:
\begin{IEEEeqnarray}{rCl}
X &\in& [0,\amp],
  \label{eq:peak}
  \\
\E{X} & \le & \alpha_1 \amp,
  \label{eq:average}
\end{IEEEeqnarray}
for some fixed parameters $\amp>0$ and $\alpha_1\in(0,1)$.  These constraints come from (eye- and skin-) safety reasons, and from limitations  (caused by non-linearities) on the optical operating regimes of  LEDs and lasers.

Due to battery limitations on the attached RF circuit and power amplifier limitations, the \emph{second moment} of the transmit signal also has  to be restricted:
\begin{equation}
\label{eq:second}
\E{X^2}  \le \alpha_2 \amp^2.
\end{equation}

We denote the capacity of the channel \eqref{eq:channel} with allowed
peak power $\amp$,  maximum  average power $\alpha_1 \amp$, and  maximum  second moment  $\alpha_2 \amp^2$  by
$\C(\alpha_1,\alpha_2,\amp)$. It is given by
\cite{Cover}
\begin{IEEEeqnarray}{c}
  \C(\alpha_1,\alpha_2,\amp)= \sup_{P_X} \II(X; Y),
  \label{eq:C}
\end{IEEEeqnarray}
where the supremum is over  input laws $P_X$  satisfying  \eqref{eq:peak}--\eqref{eq:second}.

Notice that, for any random variable $X\in[0,\amp]$, we have  $\E{X^2}\leq \E{X}\amp$ and of course $\E{X}\leq \sqrt{\E{X^2}}$. Therefore, whenever $\alpha_1 <\alpha_2$, the  second moment constraint \eqref{eq:second} is  inactive in view of the first moment-constraint \eqref{eq:average}, and whenever $\sqrt{\alpha_2} <\alpha_2$, the  first moment constraint \eqref{eq:average} is  inactive in view of the second moment-constraint \eqref{eq:second}.  Moreover,  for any $\alpha_1\geq 1/2$, the first-moment constraint \eqref{eq:average} is not active, and for  $\alpha_2\geq 1/2$, the second-moment constraint \eqref{eq:second}.  In fact, by the symmetry of the Gaussian density, for any input $X$, we have $\II(X;Y)=\II(X';Y)$ for the derived input $X'=\amp-X$, which has smaller first and second moments than $X$ if $\E{X}\geq 1/2\amp$:
\begin{equation}
\E{X'}  =\amp- \E{X} \leq 1/2 \amp \leq  \E{X},
\end{equation}
and
\begin{IEEEeqnarray}{c}
\E{X'^2}  =\amp^2- 2 \E{X}\amp + \E{X^2}  \leq  \E{X^2}.
\end{IEEEeqnarray}
We can thus limit the optimization in \eqref{eq:C} to random variables $X$ with first moments not exceeding $1/2\amp$ and, by  $\E{X^2} \leq  \E{X}\amp$, with second moments not exceeding $1/2\amp^2$.

 As a consequence:
\begin{IEEEeqnarray}{c}\label{eq:no1}
	\C(\alpha_1,\alpha_2,\amp)=	\C(1,\alpha_2,\amp), \qquad \forall \alpha_1 \geq\max\{\sqrt{\alpha_2} ,1/2\}, \IEEEeqnarraynumspace
\end{IEEEeqnarray}
and
\begin{IEEEeqnarray}{c}\label{eq:no2}
	\C(\alpha_1,\alpha_2,\amp)= 	\C(\alpha_1,1,\amp), \qquad \forall \alpha_2 \geq\max\{\alpha_1,1/2\}.\IEEEeqnarraynumspace
\end{IEEEeqnarray}

 In the remainder of the paper, we present bounds on the capacities, and establish the exact asymptotic results in the high and low SNR regimes, respectively.

 The following functions will be used throughout the paper. For $i=0,1,2,3,4$, define:
\begin{equation}\label{eq:zeta}
 \zeta_i(\lambda_1,\lambda_2):= \int_{0}^1 y^i e^{-\lambda_1 y  - \lambda_2y^2} \mathrm{d} y.
\end{equation}

\section{The asymptotic high-SNR capacity}

Consider first the asymptotic high-SNR regime, where  $\alpha_1, \alpha_2$ are fixed and  $\amp$ grows without bound. 

\begin{theorem}\label{thm:High-SNR}Depending on the parameters $\alpha_1, \alpha_2>0$, the asymptotic high-SNR capacity satisfies one of the following limiting behaviours.
\begin{enumerate}
\item If $\alpha_1 \geq \frac{1}{2}$ and $\alpha_2 \geq \frac{1}{3}$, then both the first- and second-moment  constraints are inactive and
\begin{IEEEeqnarray}{rCl}
\varlimsup_{\amp \to \infty}\left( \C(\alpha_1,\alpha_2,\amp) -\log \frac{\amp }{\sqrt{2\pi e\sigma^2}}\right)=0.
\end{IEEEeqnarray}

\item If $0<\alpha_1</2$ is such that the unique  solution $\lambda_1^*$  to the equation (in $\lambda_1$)
 \begin{subequations}
\begin{IEEEeqnarray}{rCl}
\frac{1}{\lambda_1} - \frac{e^{-\lambda_1}}{1- e^{-\lambda_1}} & = & \alpha_1
\end{IEEEeqnarray}
satisfies
\begin{IEEEeqnarray}{rCl}
\frac{2}{(\lambda_1^*)^2} - \frac{e^{-\lambda_1^*}\left(1+\frac{2}{\lambda_1^*} \right)}{1- e^{-\lambda_1^*}} & < & \alpha_2,
\end{IEEEeqnarray}
then only the first moment constraint is active and
\begin{IEEEeqnarray}{rCl}
\lefteqn{\varlimsup_{\amp \to \infty}\left( \C(\alpha_1,\alpha_2,\amp) -\log \frac{\amp }{\sqrt{2\pi e\sigma^2}}\right) } \hspace{2cm} \nonumber \\[1.2ex]
& = & \log \zeta_0(\lambda_1^*,0)+\lambda_1^* \alpha_1. \hspace{2cm}
\end{IEEEeqnarray}
\end{subequations}
 \item If $0<\alpha_2</3$ is such that the unique  solution $\lambda_2^*$  to the equation (in $\lambda_2$)
\begin{subequations}
\begin{IEEEeqnarray}{rCl}
2 \sqrt{\pi\lambda_2}  \left( (2\lambda_2)^{-1}-\alpha_2 \right) \left[ \frac{1}{2} - \Q\left( \sqrt{2 \lambda_2}\right)  \right] =  e^{- \lambda_2},\nonumber\\
\end{IEEEeqnarray}
satisfies
\begin{IEEEeqnarray}{rCl}
2\sqrt{ \pi \lambda_2^* } \alpha_1  \left[  \frac{1}{2} -\Q\left(\sqrt{2 \lambda_2^*}\right)  \right]
> 1- e^{- \lambda_2^*} \hspace{2cm}
\end{IEEEeqnarray}
then only the second moment constraint is active and
\begin{IEEEeqnarray}{rCl}
\lefteqn{\varlimsup_{\amp \to \infty}\left( \C(\alpha_1,\alpha_2,\amp) -\log \frac{\amp }{\sqrt{2\pi e\sigma^2}}\right) }\hspace{2cm} \nonumber \\[1.2ex]
& = & \log \zeta_0(0,\lambda_2^*)+\lambda_2^* \alpha_2. \hspace{2cm}
\end{IEEEeqnarray}
\end{subequations}
\item Else, both moment constraints are active and
\begin{IEEEeqnarray}{rCl}
\lefteqn{\varlimsup_{\amp \to \infty}\left( \C(\alpha_1,\alpha_2,\amp) -\log \frac{\amp }{\sqrt{2\pi e\sigma^2}}\right) }\hspace{1.2cm} \nonumber \\[1.2ex]
& = & \log \zeta_0(\lambda_1^*,\lambda_2^*)+\lambda_1^* \alpha_1 + \lambda_2^* \alpha_2,
\end{IEEEeqnarray}
for $\lambda_1^*, \lambda_2^*>0$ the unique solution to the equations
\begin{subequations}
\begin{IEEEeqnarray}{rCl}
\lefteqn{ \sqrt{ \pi \lambda_2 } e^{\frac{\lambda_1^2}{4 \lambda_2}} \left(2\alpha_1+\frac{\lambda_1}{\lambda_2}\right)   \left[ \Q\left(\frac{ \lambda_1}{\sqrt{2 \lambda_2}}\right)-\Q\left(\frac{ \lambda_1+2\lambda_2}{\sqrt{2 \lambda_2}}\right)  \right]} \hspace{3.5cm} \nonumber \\
&
=& 1- e^{- (\lambda_1+\lambda_2)} \hspace{2.1cm}
\end{IEEEeqnarray}
and
\begin{IEEEeqnarray}{rCl}
\lefteqn{ \sqrt{\frac{\pi}{ \lambda_2 }}  e^{\frac{\lambda_1^2}{4 \lambda_2}} \left(\alpha_2 - \frac{\lambda_2-\lambda_1^2}{2\lambda_2^2}\right) \left[ \Q\left(\frac{ \lambda_1}{\sqrt{2 \lambda_2}}\right)-\Q\left(\frac{ \lambda_1+2\lambda_2}{\sqrt{2 \lambda_2}}\right)  \right]}\qquad\nonumber\\
& =&  \frac{1}{2\lambda_2}e^{- (\lambda_1+\lambda_2)} \left( \frac{\lambda_1}{\lambda_2}-1\right)- \frac{\lambda_1}{2 \lambda_2^2} .\hspace{2.1cm}
\end{IEEEeqnarray}
\end{subequations}

\end{enumerate}
\end{theorem}

\section{The asymptotic low-SNR capacity}
Consider now the asymptotic low-SNR regime, where $\alpha_1, \alpha_2$ are again kept fixed and  $\amp\to 0$.
\begin{proposition}\label{prop:low}
Given parameters $\alpha_1, \alpha_2>0$,
\begin{IEEEeqnarray}{rCl}
\lim_{\amp \downarrow 0} \frac{ \C(\alpha_1,\alpha_2,\amp)}{\amp^2} = \max_{ T \in [0,1] \colon \substack{\E{T}\leq \alpha_1 \\ \frac{1}{2}\E{T^2}\leq \alpha_2} }\Var{T}.
\end{IEEEeqnarray}
\end{proposition}
\begin{IEEEproof}
The achievability follows directly from Prelov's and Verd\'u's classical result on the mutual information of peak-constrained channels \cite[Corollary 2]{prelovverdu04_1}. The converse follows by the well-known Gaussian  max-entropy bound:
\begin{equation*}
\C(\alpha_1, \alpha_2, \amp)  \leq \max \frac{1}{2} \log \left( 1+\frac{\Var{X}}{\sigma^2} \right) ,
\end{equation*}
where the maximization is over random variables $X\in [0,1]$ satisfying \eqref{eq:peak}--\eqref{eq:second}. Defining $T:=X/\amp$ and using that $\lim_{t\downarrow 0} \frac{\log(1+bt)}{t} =b$, for any constant $b>0$, establishes the  desired asymptotic converse bound.
\end{IEEEproof}
\begin{lemma}\label{lem:low}
The maximization in Proposition~\ref{prop:low} is attained by a binary random variable $T\in\{0,\amp\}$:
\begin{equation}\label{lemma:binary}
 \max_{  \substack{T \in [0,1] \colon\\ \E{T}\leq \alpha_1 \\ \E{T^2}\leq \alpha_2}} \textnormal{Var}(T)=  \max_{ \substack{ T \in \{0,\amp\} \colon \\ \E{T}\leq \alpha_1 \\ \E{T^2}\leq \alpha_2} }\textnormal{Var}(T)
 \end{equation}
\end{lemma}
\begin{IEEEproof}
Fix $T$ satisfying the conditions in the minimization and construct a new random variable $T'\in\{0,\amp\}$ with $p_{\amp}:=\Prv{T'=\amp}= \frac{ \E{T^2}}{\amp^2}$ and $\Prv{T'=0}=1-p_{\amp}$. Notice that $\E{(T')^2} = p_{\amp}\amp^2 = \E{T^2}$ and
\begin{equation}
\E{T'} =  p_{\amp}\amp = \frac{  \E{T^2}}{\amp} \leq \frac{  \E{T}\cdot \amp}{\amp} = \E{T}.
\end{equation}
The new random variable $T'$ thus also satisfies the conditions in the maximization, and moreover it has larger objective function (variance) than $T$ because $\Var{T'} = \E{(T')^2} - (\E{T'})^2\geq \E{(T)^2} - (\E{T})^2= \Var{T}$.
\end{IEEEproof}

Combining Proposition~\ref{prop:low} with Lemma~\ref{lem:low} establishes the desired low-SNR asymptotics.
\begin{theorem}For any parameters $\alpha_1,\alpha_2>0$:
\begin{equation}
\lim_{\amp \downarrow 0} \frac{ \C(\alpha_1,\alpha_2,\amp)}{\amp^2} = p^*(1-p^*),
\end{equation} where $p^*:=\min\{\alpha_1,\alpha_2, 1/2\}$.
\end{theorem}
\begin{IEEEproof}
By Lemma~\ref{lem:low}:
\begin{equation}
 \max_{  \substack{T \in [0,1] \colon\\ \E{T}\leq \alpha_1 \\ \E{T^2}\leq \alpha_2}} \textnormal{Var}(T)=  \max_{  \substack{p_{\amp} \in [0,1] \colon\\ p_{\amp} \leq \alpha_1 \\ p_{\amp}\leq  \alpha_2}} p_{\amp} (1-p_{\amp}).
 \end{equation}
 Since the function $t \mapsto t (1-t)$ is continuous and  monotonically increasing over $[0,1/2]$ but  monotonically decreasing over $[1/2,1]$, the maximum value is obtained for $p_{\amp}= \min\{\alpha_1, \alpha_2, 1/2\}$. Plugging this into Proposition~\ref{prop:low} establishes the desired result.
\end{IEEEproof}

\section{Discussion of Asymptotic Results}
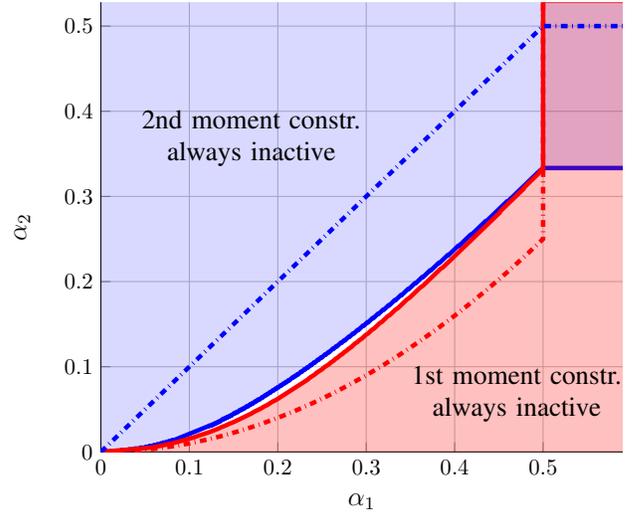
\begin{figure}
\centering
\begin{tikzpicture} [every pin/.style={fill=white},scale=0.85]
  \begin{axis}[scale=.9,
width=0.5\textwidth,
scale only axis,
xmin=0,
xmax=0.59,
xmajorgrids,
xlabel={\large{$\alpha_1$}},
ymin=0,
ymax=0.528,
ymajorgrids,
ylabel={\large{$\alpha_2$}},
axis x line*=bottom,
axis y line*=left,
legend pos=north west,
legend style={draw=none,fill=none,legend cell align=left, font=\normalsize}
]

   \addplot[color=red,dashdotted,line width=2pt]
        	table[row sep=crcr]{
        		0.00000   0.00000\\
        		0.02000   0.00040\\
        		0.04000   0.00160\\
        		0.06000   0.00360\\
        		0.08000   0.00640\\
        		0.10000   0.01000\\
        		0.12000   0.01440\\
        		0.14000   0.0196\\
        		0.16000   0.02560\\
        		0.18000   0.03240\\
        		0.20000   0.04000\\
        		0.22000   0.04840\\
        		0.24000   0.05760\\
        		0.26000   0.06760\\
        		0.28000   0.07840\\
        		0.30000   0.09000\\
        		0.32000   0.10240\\
        		0.34000   0.11560\\
        		0.36000   0.12960\\
        		0.38000   0.14440\\
        		0.40000   0.16000\\
        		0.42000   0.17640\\
        		0.44000   0.19360\\
        		0.46000   0.21160\\
        		0.48000   0.23040\\
        		0.50000   0.25000\\
%
0.50000000   0.54\\
0.6 0.54\\
0.6 0\\};

         \addplot[color=blue,solid,line width=2pt, fill=blue,
                    fill opacity=0.15]
 table[row sep=crcr]{
 0 0\\
    0.00020000   0.00020000\\
   0.00700000   0.00020000\\
   0.00800000   0.00050000\\
   0.01500000   0.00050000\\
   0.01600000   0.00080000\\
   0.01900000   0.00080000\\
   0.02000000   0.00100000\\
   0.02200000   0.00100000\\
   0.02300000   0.00200000\\
   0.03100000   0.00200000\\
   0.03200000   0.00300000\\
   0.03800000   0.00300000\\
   0.03900000   0.00400000\\
   0.04400000   0.00400000\\
   0.04500000   0.00500000\\
   0.05000000   0.00500000\\
   0.05100000   0.00600000\\
   0.05400000   0.00600000\\
   0.05500000   0.00700000\\
   0.05900000   0.00700000\\
   0.06000000   0.00800000\\
   0.06300000   0.00800000\\
   0.06400000   0.00900000\\
   0.06700000   0.00900000\\
   0.06800000   0.01000000\\
   0.07000000   0.01000000\\
   0.07100000   0.01100000\\
   0.07400000   0.01100000\\
   0.07500000   0.01200000\\
   0.07600000   0.01200000\\
   0.07700000   0.01200000\\
   0.07800000   0.01300000\\
   0.08000000   0.01300000\\
   0.08100000   0.01400000\\
   0.08300000   0.01400000\\
   0.12100000   0.03000000\\
   0.12200000   0.03000000\\
   0.12300000   0.03100000\\
   0.12400000   0.03100000\\
   0.12500000   0.03100000\\
   0.12600000   0.03200000\\
   0.12700000   0.03200000\\
   0.12800000   0.03300000\\
   0.12900000   0.03300000\\
   0.13000000   0.03400000\\
   0.13100000   0.03400000\\
   0.13200000   0.03500000\\
   0.13300000   0.03500000\\
   0.13400000   0.03600000\\
   0.13500000   0.03600000\\
   0.13600000   0.03700000\\
   0.13700000   0.03800000\\
   0.13800000   0.03800000\\
   0.13900000   0.03900000\\
   0.14000000   0.03900000\\
   0.14100000   0.04000000\\
   0.14200000   0.04000000\\
   0.14300000   0.04100000\\
   0.14400000   0.04100000\\
   0.14500000   0.04200000\\
   0.14600000   0.04200000\\
   0.14700000   0.04300000\\
   0.14800000   0.04300000\\
   0.14900000   0.04400000\\
   0.15000000   0.04500000\\
   0.15100000   0.04500000\\
   0.15200000   0.04600000\\
   0.15300000   0.04600000\\
   0.15400000   0.04700000\\
   0.15500000   0.04700000\\
   0.15600000   0.04800000\\
   0.15700000   0.04900000\\
   0.15800000   0.04900000\\
   0.15900000   0.05000000\\
   0.16000000   0.05000000\\
   0.16100000   0.05100000\\
   0.16200000   0.05100000\\
   0.16300000   0.05200000\\
   0.16400000   0.05300000\\
   0.16500000   0.05300000\\
   0.16600000   0.05400000\\
   0.16700000   0.05400000\\
   0.16800000   0.05500000\\
   0.16900000   0.05600000\\
   0.17000000   0.05600000\\
   0.17100000   0.05700000\\
   0.17200000   0.05700000\\
   0.17300000   0.05800000\\
   0.17400000   0.05900000\\
   0.17500000   0.05900000\\
   0.17600000   0.06000000\\
   0.17700000   0.06100000\\
   0.17800000   0.06100000\\
   0.17900000   0.06200000\\
   0.18000000   0.06200000\\
   0.18100000   0.06300000\\
   0.18200000   0.06400000\\
   0.18300000   0.06400000\\
   0.18400000   0.06500000\\
   0.18500000   0.06600000\\
   0.18600000   0.06600000\\
   0.18700000   0.06700000\\
   0.18800000   0.06800000\\
   0.18900000   0.06800000\\
   0.19000000   0.06900000\\
   0.19100000   0.07000000\\
   0.19200000   0.07000000\\
   0.19300000   0.07100000\\
   0.19400000   0.07200000\\
   0.19500000   0.07200000\\
   0.19600000   0.07300000\\
   0.19700000   0.07400000\\
   0.19800000   0.07400000\\
   0.19900000   0.07500000\\
   0.20000000   0.07600000\\
   0.20100000   0.07600000\\
   0.20200000   0.07700000\\
   0.20300000   0.07800000\\
   0.20400000   0.07800000\\
   0.20500000   0.07900000\\
   0.20600000   0.08000000\\
   0.20700000   0.08000000\\
   0.20800000   0.08100000\\
   0.20900000   0.08200000\\
   0.21000000   0.08200000\\
   0.21100000   0.08300000\\
   0.21200000   0.08400000\\
   0.21300000   0.08400000\\
   0.21400000   0.08500000\\
   0.21500000   0.08600000\\
   0.21600000   0.08700000\\
   0.21700000   0.08700000\\
   0.21800000   0.08800000\\
   0.21900000   0.08900000\\
   0.22000000   0.08900000\\
   0.22100000   0.09000000\\
   0.22200000   0.09100000\\
   0.22300000   0.09100000\\
   0.22400000   0.09200000\\
   0.22500000   0.09300000\\
   0.22600000   0.09400000\\
   0.22700000   0.09400000\\
   0.22800000   0.09500000\\
   0.22900000   0.09600000\\
   0.23000000   0.09700000\\
   0.23100000   0.09700000\\
   0.23200000   0.09800000\\
   0.23300000   0.09900000\\
   0.23400000   0.09900000\\
   0.23500000   0.10000000\\
   0.23600000   0.10100000\\
   0.23700000   0.10200000\\
   0.23800000   0.10200000\\
   0.23900000   0.10300000\\
   0.24000000   0.10400000\\
   0.24100000   0.10500000\\
   0.24200000   0.10500000\\
   0.24300000   0.10600000\\
   0.24400000   0.10700000\\
   0.24500000   0.10800000\\
   0.24600000   0.10800000\\
   0.24700000   0.10900000\\
   0.24800000   0.11000000\\
   0.24900000   0.11100000\\
   0.25000000   0.11100000\\
   0.25100000   0.11200000\\
   0.25200000   0.11300000\\
   0.25300000   0.11400000\\
   0.25400000   0.11400000\\
   0.25500000   0.11500000\\
   0.25600000   0.11600000\\
   0.25700000   0.11700000\\
   0.25800000   0.11700000\\
   0.25900000   0.11800000\\
   0.26000000   0.11900000\\
   0.26100000   0.12000000\\
   0.26200000   0.12100000\\
   0.26300000   0.12100000\\
   0.26400000   0.12200000\\
   0.26500000   0.12300000\\
   0.26600000   0.12400000\\
   0.26700000   0.12400000\\
   0.26800000   0.12500000\\
   0.26900000   0.12600000\\
   0.27000000   0.12700000\\
   0.27100000   0.12800000\\
   0.27200000   0.12800000\\
   0.27300000   0.12900000\\
   0.27400000   0.13000000\\
   0.27500000   0.13100000\\
   0.27600000   0.13100000\\
   0.27700000   0.13200000\\
   0.27800000   0.13300000\\
   0.27900000   0.13400000\\
   0.28000000   0.13500000\\
   0.28100000   0.13500000\\
   0.28200000   0.13600000\\
   0.28300000   0.13700000\\
   0.28400000   0.13800000\\
   0.28500000   0.13900000\\
   0.28600000   0.13900000\\
   0.28700000   0.14000000\\
   0.28800000   0.14100000\\
   0.28900000   0.14200000\\
   0.29000000   0.14300000\\
   0.29100000   0.14300000\\
   0.29200000   0.14400000\\
   0.29300000   0.14500000\\
   0.29400000   0.14600000\\
   0.29500000   0.14700000\\
   0.29600000   0.14800000\\
   0.29700000   0.14800000\\
   0.29800000   0.14900000\\
   0.29900000   0.15000000\\
   0.30000000   0.15100000\\
   0.30100000   0.15200000\\
   0.30200000   0.15200000\\
   0.30300000   0.15300000\\
   0.30400000   0.15400000\\
   0.30500000   0.15500000\\
   0.30600000   0.15600000\\
   0.30700000   0.15700000\\
   0.30800000   0.15700000\\
   0.30900000   0.15800000\\
   0.31000000   0.15900000\\
   0.31100000   0.16000000\\
   0.31200000   0.16100000\\
   0.31300000   0.16200000\\
   0.31400000   0.16200000\\
   0.31500000   0.16300000\\
   0.31600000   0.16400000\\
   0.31700000   0.16500000\\
   0.31800000   0.16600000\\
   0.31900000   0.16700000\\
   0.32000000   0.16700000\\
   0.32100000   0.16800000\\
   0.32200000   0.16900000\\
   0.32300000   0.17000000\\
   0.32400000   0.17100000\\
   0.32500000   0.17200000\\
   0.32600000   0.17200000\\
   0.32700000   0.17300000\\
   0.32800000   0.17400000\\
   0.32900000   0.17500000\\
   0.33000000   0.17600000\\
   0.33100000   0.17700000\\
   0.33200000   0.17800000\\
   0.33300000   0.17800000\\
   0.33400000   0.17900000\\
   0.33500000   0.18000000\\
   0.33600000   0.18100000\\
   0.33700000   0.18200000\\
   0.33800000   0.18300000\\
   0.33900000   0.18400000\\
   0.34000000   0.18400000\\
   0.34100000   0.18500000\\
   0.34200000   0.18600000\\
   0.34300000   0.18700000\\
   0.34400000   0.18800000\\
   0.34500000   0.18900000\\
   0.34600000   0.19000000\\
   0.34700000   0.19100000\\
   0.34800000   0.19100000\\
   0.34900000   0.19200000\\
   0.35000000   0.19300000\\
   0.35100000   0.19400000\\
   0.35200000   0.19500000\\
   0.35300000   0.19600000\\
   0.35400000   0.19700000\\
   0.35500000   0.19800000\\
   0.35600000   0.19800000\\
   0.35700000   0.19900000\\
   0.35800000   0.20000000\\
   0.35900000   0.20100000\\
   0.36000000   0.20200000\\
   0.36100000   0.20300000\\
   0.36200000   0.20400000\\
   0.36300000   0.20500000\\
   0.36400000   0.20500000\\
   0.36500000   0.20600000\\
   0.36600000   0.20700000\\
   0.36700000   0.20800000\\
   0.36800000   0.20900000\\
   0.36900000   0.21000000\\
   0.37000000   0.21100000\\
   0.37100000   0.21200000\\
   0.37200000   0.21300000\\
   0.37300000   0.21300000\\
   0.37400000   0.21400000\\
   0.37500000   0.21500000\\
   0.37600000   0.21600000\\
   0.37700000   0.21700000\\
   0.37800000   0.21800000\\
   0.37900000   0.21900000\\
   0.38000000   0.22000000\\
   0.38100000   0.22100000\\
   0.38200000   0.22200000\\
   0.38300000   0.22200000\\
   0.38400000   0.22300000\\
   0.38500000   0.22400000\\
   0.38600000   0.22500000\\
   0.38700000   0.22600000\\
   0.38800000   0.22700000\\
   0.38900000   0.22800000\\
   0.39000000   0.22900000\\
   0.39100000   0.23000000\\
   0.39200000   0.23100000\\
   0.39300000   0.23100000\\
   0.39400000   0.23200000\\
   0.39500000   0.23300000\\
   0.39600000   0.23400000\\
   0.39700000   0.23500000\\
   0.39800000   0.23600000\\
   0.39900000   0.23700000\\
   0.40000000   0.23800000\\
   0.40100000   0.23900000\\
   0.40200000   0.24000000\\
   0.40300000   0.24100000\\
   0.40400000   0.24200000\\
   0.40500000   0.24200000\\
   0.40600000   0.24300000\\
   0.40700000   0.24400000\\
   0.40800000   0.24500000\\
   0.40900000   0.24600000\\
   0.41000000   0.24700000\\
   0.41100000   0.24800000\\
   0.41200000   0.24900000\\
   0.41300000   0.25000000\\
   0.41400000   0.25100000\\
   0.41500000   0.25200000\\
   0.41600000   0.25300000\\
   0.41700000   0.25400000\\
   0.41800000   0.25500000\\
   0.41900000   0.25500000\\
   0.42000000   0.25600000\\
   0.42100000   0.25700000\\
   0.42200000   0.25800000\\
   0.42300000   0.25900000\\
   0.42400000   0.26000000\\
   0.42500000   0.26100000\\
   0.42600000   0.26200000\\
   0.42700000   0.26300000\\
   0.42800000   0.26400000\\
   0.42900000   0.26500000\\
   0.43000000   0.26600000\\
   0.43100000   0.26700000\\
   0.43200000   0.26800000\\
   0.43300000   0.26900000\\
   0.43400000   0.27000000\\
   0.43500000   0.27100000\\
   0.43600000   0.27100000\\
   0.43700000   0.27200000\\
   0.43800000   0.27300000\\
   0.43900000   0.27400000\\
   0.44000000   0.27500000\\
   0.44100000   0.27600000\\
   0.44200000   0.27700000\\
   0.44300000   0.27800000\\
   0.44400000   0.27900000\\
   0.44500000   0.28000000\\
   0.44600000   0.28100000\\
   0.44700000   0.2820000\\
   0.44800000   0.28300000\\
   0.44900000   0.28400000\\
   0.45000000   0.28500000\\
   0.45100000   0.28600000\\
   0.45200000   0.28700000\\
   0.45300000   0.28800000\\
   0.45400000   0.28900000\\
   0.45500000   0.29000000\\
   0.45600000   0.29100000\\
   0.45700000   0.29200000\\
   0.45800000   0.29300000\\
   0.45900000   0.29400000\\
   0.46000000   0.29400000\\
   0.46100000   0.29500000\\
   0.46200000   0.29600000\\
   0.46300000   0.29700000\\
   0.46400000   0.29800000\\
   0.46500000   0.29900000\\
   0.46600000   0.30000000\\
   0.46700000   0.30100000\\
   0.46800000   0.30200000\\
   0.46900000   0.30300000\\
   0.47000000   0.30400000\\
   0.47100000   0.30500000\\
   0.47200000   0.30600000\\
   0.47300000   0.30700000\\
   0.47400000   0.30800000\\
   0.47500000   0.30900000\\
   0.47600000   0.31000000\\
   0.47700000   0.31100000\\
   0.47800000   0.31200000\\
   0.47900000   0.31300000\\
   0.48000000   0.31400000\\
   0.48100000   0.31500000\\
   0.48200000   0.31600000\\
   0.48300000   0.31700000\\
   0.48400000   0.31800000\\
   0.48500000   0.31900000\\
   0.48600000   0.32000000\\
   0.48700000   0.32100000\\
   0.48800000   0.32200000\\
   0.48900000   0.32300000\\
   0.49000000   0.32400000\\
   0.49100000   0.32500000\\
   0.49200000   0.32600000\\
   0.49300000   0.32700000\\
   0.49400000   0.32800000\\
   0.49500000   0.32900000\\
   0.49600000   0.33000000\\
      0.50000000   0.3333333\\
   0.60000000   0.3333333\\
   0.6 0.54\\
   0 0.54\\};

 \addplot[color=red, solid,line width=2pt, fill=red,
                    fill opacity=0.25]
 table[row sep=crcr]{
   0 0\\
   0.05700000   0.00500000\\
   0.07600000   0.00900000\\
   0.08400000   0.01100000\\
   0.09100000   0.01300000\\
   0.09800000   0.01500000\\
   0.10500000   0.01700000\\
   0.11000000   0.01900000\\
   0.11600000   0.02100000\\
   0.12200000   0.02300000\\
   0.12700000   0.02500000\\
   0.13200000   0.02700000\\
   0.13600000   0.02900000\\
   0.14100000   0.03100000\\
   0.14500000   0.03300000\\
   0.15000000   0.03500000\\
   0.15400000   0.03700000\\
   0.15800000   0.03900000\\
   0.16200000   0.04100000\\
   0.16600000   0.04300000\\
   0.17000000   0.04500000\\
   0.17300000   0.04700000\\
   0.17700000   0.04900000\\
   0.18100000   0.05100000\\
   0.18400000   0.05300000\\
   0.18800000   0.05500000\\
   0.19100000   0.05700000\\
   0.19400000   0.05900000\\
   0.19800000   0.06100000\\
   0.20100000   0.06300000\\
   0.20400000   0.06500000\\
   0.20700000   0.06700000\\
   0.21000000   0.06900000\\
   0.21300000   0.07100000\\
   0.21600000   0.07300000\\
   0.21900000   0.07500000\\
   0.22200000   0.07700000\\
   0.22500000   0.07900000\\
   0.22800000   0.08100000\\
   0.23100000   0.08300000\\
   0.23400000   0.08500000\\
   0.23600000   0.08700000\\
   0.23900000   0.08900000\\
   0.24200000   0.09100000\\
   0.24400000   0.09300000\\
   0.24700000   0.09500000\\
   0.25000000   0.09700000\\
   0.25200000   0.09900000\\
   0.25500000   0.10100000\\
   0.25800000   0.10300000\\
   0.26000000   0.10500000\\
   0.26300000   0.10700000\\
   0.26500000   0.10900000\\
   0.26800000   0.11100000\\
   0.27000000   0.11300000\\
   0.27300000   0.11500000\\
   0.27500000   0.11700000\\
   0.27800000   0.11900000\\
   0.28000000   0.12100000\\
   0.28300000   0.12300000\\
   0.28500000   0.12500000\\
   0.28700000   0.12700000\\
   0.29000000   0.12900000\\
   0.29200000   0.13100000\\
   0.29500000   0.13300000\\
   0.29700000   0.13500000\\
   0.29900000   0.13700000\\
   0.30200000   0.13900000\\
   0.30400000   0.14100000\\
   0.30600000   0.14300000\\
   0.30900000   0.14500000\\
   0.31100000   0.14700000\\
   0.31300000   0.14900000\\
   0.31500000   0.15100000\\
   0.31800000   0.15300000\\
   0.32000000   0.15500000\\
   0.32200000   0.15700000\\
   0.32400000   0.15900000\\
   0.32700000   0.16100000\\
   0.32900000   0.16300000\\
   0.33100000   0.16500000\\
   0.33300000   0.16700000\\
   0.33600000   0.16900000\\
   0.33800000   0.17100000\\
   0.34000000   0.17300000\\
   0.34200000   0.17500000\\
   0.34400000   0.17700000\\
   0.34600000   0.17900000\\
   0.34900000   0.18100000\\
   0.35100000   0.18300000\\
   0.35300000   0.18500000\\
   0.35500000   0.18700000\\
   0.35700000   0.18900000\\
   0.35900000   0.19100000\\
   0.36100000   0.19300000\\
   0.36400000   0.19500000\\
   0.36600000   0.19700000\\
   0.36800000   0.19900000\\
   0.37000000   0.20100000\\
   0.37200000   0.20300000\\
   0.37400000   0.20500000\\
   0.37600000   0.20700000\\
   0.37800000   0.20900000\\
   0.38000000   0.21100000\\
   0.38200000   0.21300000\\
   0.38400000   0.21500000\\
   0.38700000   0.21700000\\
   0.38900000   0.21900000\\
   0.39100000   0.22100000\\
   0.39300000   0.22300000\\
   0.39500000   0.22500000\\
   0.39700000   0.22700000\\
   0.39900000   0.22900000\\
   0.40100000   0.23100000\\
   0.40300000   0.23300000\\
   0.40500000   0.23500000\\
   0.40700000   0.23700000\\
   0.40900000   0.23900000\\
   0.41100000   0.24100000\\
   0.41300000   0.24300000\\
   0.41500000   0.24500000\\
   0.41700000   0.24700000\\
   0.41900000   0.24900000\\
   0.42100000   0.25100000\\
   0.42300000   0.25300000\\
   0.42500000   0.25500000\\
   0.42700000   0.25700000\\
   0.42900000   0.25900000\\
   0.43100000   0.26100000\\
   0.43300000   0.26300000\\
   0.43500000   0.26500000\\
   0.43700000   0.26700000\\
   0.43900000   0.26900000\\
   0.44100000   0.27100000\\
   0.44300000   0.27300000\\
   0.44500000   0.27500000\\
   0.44700000   0.27700000\\
   0.44900000   0.27900000\\
   0.45100000   0.28100000\\
   0.45200000   0.28300000\\
   0.45400000   0.28500000\\
   0.45600000   0.28700000\\
   0.45800000   0.28900000\\
   0.46000000   0.29100000\\
   0.46200000   0.29300000\\
   0.46400000   0.29500000\\
   0.46600000   0.29700000\\
   0.46800000   0.29900000\\
   0.47000000   0.30100000\\
   0.47200000   0.30300000\\
   0.47400000   0.30500000\\
   0.47600000   0.30700000\\
   0.47800000   0.30900000\\
   0.47900000   0.31100000\\
   0.48100000   0.31300000\\
   0.48300000   0.31500000\\
   0.48500000   0.31700000\\
   0.48700000   0.31900000\\
   0.48900000   0.32100000\\
   0.49100000   0.32300000\\
   0.49300000   0.32500000\\
   0.49500000   0.32700000\\
   0.49600000   0.32900000\\
0.5 0.33333\\
0.5 0.53 \\
0.63 0.53\\
0.63 0\\};

   \addplot[color=blue,dashdotted,line width=2pt]
 table[row sep=crcr]{
 0 0 \\
 0.5 0.5 \\
 0.6 0.5\\
 0.6 0.54\\
 0 0.54\\};


    \node [rotate=0] at (axis cs:  .17,  .39) {\large 2nd moment constr.};
    \node [rotate=0] at (axis cs:  .17,  .35) {\large always inactive};

        \node [rotate=0] at (axis cs:  .47,  .09) {\large 1st moment constr.};
        \node [rotate=0] at (axis cs:  .47,  .05) {\large always inactive};
\end{axis}

\end{tikzpicture}
\caption{The figure illustrates the regions where the two moment constraints \eqref{eq:average} and \eqref{eq:second} limit the (asymptotic) capacity.}
\vspace{-2mm}\label{fig:alpha12_plot}

\end{figure}

Figure~\ref{fig:alpha12_plot} illustrates the regions of $(\alpha_1, \alpha_2)$-pairs where both the first and the second-moment constraints, i.e., \eqref{eq:average} and \eqref{eq:second}, are active. At any SNR values, the first-moment constraint \eqref{eq:average} is not active on the right of the the blue dash-dotted  line, see \eqref{eq:no1}, and the second-moment constraint \eqref{eq:average} is not active above the red dash-dotted  line, see \eqref{eq:no2}.

In the asymptotic low-SNR regime, only one of the two constraints is active, unless $\alpha_1=\alpha_2< 1/2$ in which case both constraints are active, or $\alpha_1, \alpha_2 \geq 1/2$ in which case no constraint is active. Otherwise, the first-moment constraint is active when $\alpha_1<\min\{\alpha_2,1/2\}$, i.e., above the dashdotted blue line,  and the second-moment constraint is active when $\alpha_2 < \min\{\alpha_1,1/2\}$, i.e., below the dashdotted blue line.

This contrasts the high-SNR regime where both constraints are inactive for $\alpha_1\geq1/2$ and $\alpha_2\geq1/3$. Generally, the first-moment constraint \eqref{eq:average} is inactive for all $(\alpha_1, \alpha_2)$-pairs on the right of the  red solid line shown in Figure~\ref{fig:alpha12_plot}. The  second-moment constraint \eqref{eq:second} is inactive for all pairs lying above the blue solid  line. Both the first and second-moment constraints are thus simultaneously active only in the small white region that lies in between   the solid blue   and   red  lines.

Overall, it can be noted that the second-moment constraint is more stringent in the low-SNR regime than in the high-SNR regime, where it is inactive for more $(\alpha_1, \alpha_2)$-pairs. Surprisingly, we observe that in the asymptotic regimes  both constraints are simultaneously active in very few cases.

\section{Proof of Theorem~\ref{thm:High-SNR}} \label{sec:proof-HighSNR}
\subsection{Lower Bound}
We first lower-bound the capacity with some simple entropy-manipulations and by using the entropy-maximizing input-density $f_X^*(x)$ over $[0,\amp]$. Under constraints \eqref{eq:peak}--\eqref{eq:second}, $f_X^*(x)$ has  the  form:
\begin{equation}\label{eq:fX_star}
f_X^*(x) = (\amp \zeta_0(\lambda_1,\lambda_2) )^{-1}\cdot e^{- \frac{\lambda_1 }{\amp} x -\frac{ \lambda_2}{\amp^2} x^2},  \quad x \in [0,\amp],
\end{equation}
where
  the parameters $\lambda_1,\lambda_2$  have to be chosen to satisfy
	\begin{subequations}\label{eq:m}
		\begin{IEEEeqnarray}{rCl}
			\int_{0}^{\amp} f_X^*(x) \cdot x \  \mathrm{d} x & \leq   & \alpha_1 \amp, \label{eq:m1}\\
			\int_{0}^{\amp} f_X^*(x) \cdot	 x^2 \ \mathrm{d} x & \leq  & \alpha_2 \amp^2. \label{eq:m12}
		\end{IEEEeqnarray}
	\end{subequations}

Given the form in \eqref{eq:fX_star},  through a simple variable substitution $y=\frac{x}{\amp}$, one can prove that \eqref{eq:m} are equivalent to
\begin{subequations}\label{eq:m2}
\begin{IEEEeqnarray}{rCl}
   \frac{\zeta_1(\lambda_1,\lambda_2)}{   \zeta_0(\lambda_1,\lambda_2)} 	& \leq& \alpha_1 , \IEEEeqnarraynumspace\\
  \frac{\zeta_2(\lambda_1,\lambda_2)}{   \zeta_0(\lambda_1,\lambda_2)}  &\leq  &\alpha_2,
\end{IEEEeqnarray}
\end{subequations}
where recall that the functions $\zeta_i$, for $i=0,1,\ldots,4$, are defined in \eqref{eq:zeta}.
Then,
\begin{IEEEeqnarray}{rCl}
	\lefteqn{ \C(\alpha_1,\alpha_2,\amp)} \quad \nonumber \\ & \geq & 	I_{f_X^*}(X;Y) \\
	  & = & h_{f_X^*}(Y) - h(Z) \geq h_{f_X^*}(Y|Z) - h(Z) \\
	  &= &  h_{f_X^*}(X) - h(Z) \\
	   & = & \E[f_X^*]{- \log f_X^*(X)} - \frac{1}{2} \log (2 \pi e\sigma^2) \\
	   & = & \log \left(\amp \cdot \zeta_0(\lambda_1,\lambda_2)\right)  +  \frac{\lambda_1}{\amp} \E[f_X^*]{X} + \frac{\lambda_2}{\amp^2} \E[f_X^*]{X^2} \nonumber \\
	    && - \frac{1}{2} \log (2 \pi e\sigma^2) \\
 & = & \log\left(\frac{ \amp \cdot \zeta_0(\lambda_1,\lambda_2) }{\sqrt{2 \pi e \sigma^2}}\right)+ \lambda_1\alpha_1 + \lambda_2\alpha_2 , \; \IEEEeqnarraynumspace
\end{IEEEeqnarray}
where 
 all $(\lambda_1, \lambda_2)$ satisfying \eqref{eq:m2} yield valid lower bounds.

\subsection{Upper bound}
We turn to the duality-based upper bound with the choice of output density
\begin{equation}\label{eq:fy}
f_Y(y) = \tau \cdot f_Y^{(1)}(y) + (1-\tau) \cdot  f_Y^{(2)}(y),
\end{equation}
where $\tau\in(0,1)$ is a parameter that we specify later on; $f_Y^{(1)}(y)$ is a probability density function over the interval $\mathcal{I}:=[0,\amp]$ of the form
\begin{equation}\label{eq:f1}
f_Y^{(1)}(y) =\frac{1}{ \amp\cdot  \zeta_0( \lambda_1,\lambda_2)}e^{- \frac{\lambda_1}{\amp} y - \frac{\lambda_2 }{\amp} y^2} \cdot \mathbbm{1} \{ y\in\mathcal{I}\},
\end{equation}
where $\lambda_1,\lambda_2\geq0$ are free parameters, over which we will optimize in a latter stage;  and   $f_Y^{(2)}(y)$ is a probability density function over the rest of the real line $\mathcal{I}^{c}:=\mathbb{R}\backslash \mathcal{I}$:
\begin{equation}\label{eq:f2}
f_Y^{(2)}(y) = \begin{cases} \frac{1}{\sqrt{2 \pi \sigma^2}}  e^{-\frac{y^2}{2 \sigma^2}} & \textnormal{ if } y <0 ,\\
 \frac{1}{\sqrt{2 \pi \sigma^2}}  e^{-\frac{(y-\amp)^2}{2 \sigma^2}} & \textnormal{ if } y > \amp.
 \end{cases}
\end{equation}

For the choice in \eqref{eq:fy}, the duality-based upper bound yields
\begin{IEEEeqnarray}{rCl}
	\lefteqn{ \C(\alpha_1,\alpha_2,\amp)} \\ & \leq & 	\E[f_Y^*]{ - \log f_Y(Y)}  - \frac{1}{2} \log (2 \pi e\sigma^2) \\
	 & \leq & 	\E[f_Y^*]{ - \log \left(\tau f_Y^{(1)}(Y) \right) \Big|  Y\in \mathcal{I}} \cdot P_{f_Y^*}(\mathcal{I}) \nonumber \\
	 & &  + \E[f_Y^*]{ - \log \Big((1-\tau) f_Y^{(2)}(Y) \Big)\Big|  Y\in \mathcal{I}^c}\cdot  P_{f_Y^*}(\mathcal{I}^c)  \nonumber \\
	 & & - \frac{1}{2} \log (2 \pi e\sigma^2) \\
	 & = & \log \frac{ \amp \cdot \zeta_0(\lambda_1,\lambda_2) }{\tau}  \cdot   P_{f_Y^*}(\mathcal{I})\nonumber \\
	  && +    \E[f_Y^*]{ - \log \left( \tau \cdot f_Y^{(2)}(Y) \right)\Big|  Y\in \mathcal{I}^c}\cdot  P_{f_Y^*}(\mathcal{I}^c) \nonumber \\
	 & & +\left(\frac{ \lambda_1 }{\amp}\E[f_Y^*]{Y| Y \in\mathcal{I}}+ \frac{\lambda_2 }{\amp^2}\E[f_Y^*]{Y^2|Y \in \mathcal{I}} \right) \cdot P_{f_Y^*}(\mathcal{I})  \nonumber \\
& & - \frac{1}{2} \log (2 \pi e\sigma^2).\label{eq:first_step}
\end{IEEEeqnarray}
Following similar  steps as, e.g., in \cite[Eq. (209)--(226)]{Li_MIMO}, we obtain the following lemmas.
\begin{lemma}For the Gaussian-tail distribution defined in \eqref{eq:f2}:
\begin{IEEEeqnarray}{rCl}
  \E[f_Y^*]{ - \log \left( f_Y^{(2)}(Y) \right)\Big|  Y\in \mathcal{I}^c}  \leq \log \sqrt{2\pi e\sigma^2} .
  \end{IEEEeqnarray}
\end{lemma}
\begin{IEEEproof}We have
 \begin{IEEEeqnarray}{rCl}
\lefteqn{ \int_{-\infty}^{0}
  \frac{1}{\sqrt{2\pi}}
  \ope^{-\frac{(y-x)^2}{2}}
  \left(\log{\sqrt{2\pi}} +
    \frac{y^2}{2}\right) \dd y  }
  \nonumber
  \\
  & = & \log{\sqrt{2\pi}}
  \Qf{x} + \frac{1}{2}x^2
  \Qf{x}
  + \frac{1}{2}\Qf{x}
  - \frac{x}{2}
 \frac{1}{\sqrt{2\pi}}
  \ope^{-\frac{x^2}{2}}
  \IEEEeqnarraynumspace
  \\
  & \leq &  \left( \log{\sqrt{2\pi}}
  + \frac{1}{2}\right)
  \Qf{x}
  \label{eq:e207}
\end{IEEEeqnarray}
ans similarly
 \begin{IEEEeqnarray}{rCl}
\lefteqn{ \int_{\amp}^{\infty}
  \frac{1}{\sqrt{2\pi}}
  \ope^{-\frac{(y-x)^2}{2}}
  \left(\log{\sqrt{2\pi}} +
    \frac{(y-\amp)^2}{2}\right) \dd y  }
  \nonumber
  \\
  & \leq &  \left( \log{\sqrt{2\pi}}
  + \frac{1}{2}\right)
  \Qf{\amp-x}.
\end{IEEEeqnarray}
Therefore,
\begin{IEEEeqnarray}{rCl}
\lefteqn{\E[f_Y^*]{ - \log \left( f_Y^{(2)}(Y) \right)\Big|  Y\in \mathcal{I}^c}  \cdot P_{f_Y^*}(\mathcal{I}^c)  } \quad \nonumber \\
&= & - \int_{\mathcal{I}^c} f_Y^*(y) \log \left( f_Y^{(2)}(Y) \right) \dd y \\
&= &- \int_{\mathcal{I}^c} \int_{0}^{\amp}   f_X^*(x)   \frac{1}{\sqrt{2\pi}}
  \ope^{-\frac{(y-x)^2}{2}}   \dd x \cdot  \log \left( f_Y^{(2)}(Y) \right) \dd y\nonumber\\  \\
& = &- \E[f_X^*]{  \int_{\mathcal{I}^c}  \frac{1}{\sqrt{2\pi}}
  \ope^{-\frac{(y-X)^2}{2}}  \log \left( f_Y^{(2)}(Y) \right)   \dd y }\\
& = &  \left( \log{\sqrt{2\pi}e} + \frac{1}{2}\right) \E[f_X^*]{\Qf{X}+   \Qf{\amp-X} }.
  \end{IEEEeqnarray}
  Since
  \begin{IEEEeqnarray}{rCl}
   P_{f_Y^*}(\mathcal{I}^c)  & = & \E[f_X^*]{  \int_{\mathcal{I}^c}  \frac{1}{\sqrt{2\pi}}
  \ope^{-\frac{(y-X)^2}{2}}  \dd y }\\[1.2ex]
  &=& \E[f_X^*]{ \Qf{X}+   \Qf{\amp-X} }
    \end{IEEEeqnarray}
    we obtain the desired result.
\end{IEEEproof}
\begin{lemma}
For the distribution in \eqref{eq:f1}:
\begin{IEEEeqnarray}{rCl}\label{eq:proof1}
\E[f_Y^*]{Y| Y \in\mathcal{I}}  \cdot P_{f_Y^*}(\mathcal{I}) &  \leq & \E[f_Y^*]{Y} + \frac{1}{\sqrt{2\pi}} \left(1 - e^{-\frac{\amp^2}{2} }  \right) \\
& = & \E[f_X^*]{X} + \left(1 - \frac{1}{\sqrt{2\pi}} e^{-\frac{\amp^2}{2} }  \right)   \IEEEeqnarraynumspace
\end{IEEEeqnarray}
and  \begin{IEEEeqnarray}{rCl}\label{eq:proof2}
\E[f_Y^*]{Y^2|Y \in \mathcal{I}} \cdot P_{f_Y^*}(\mathcal{I})  & \leq & \E[f_Y^*]{Y^2}=  \E[f_X^*]{X^2} +\sigma^2.\IEEEeqnarraynumspace
\end{IEEEeqnarray}
\end{lemma}
\begin{IEEEproof}The inequality in  \eqref{eq:proof2} follows simply because  $Y^2 \geq 0$ with probability $1$. The inequality in  \eqref{eq:proof1} is proved as follows:
\begin{IEEEeqnarray}{rCl}
\lefteqn{\E[f_Y^*]{Y| Y \in\mathcal{I}}  \cdot P_{f_Y^*}(\mathcal{I})  } \quad \nonumber \\
&= &  \int_{0}^{\amp} f_Y^*(y) \cdot y \dd y \\
&= &  \int_{0}^{\amp} \int_{0}^{\amp}   f_X^*(x)   \frac{1}{\sqrt{2\pi}}
  \ope^{-\frac{(y-x)^2}{2}}   \dd x \cdot y \dd y  \\
& = & \E[f_X^*]{ \int_{0}^{\amp}
  \frac{1}{\sqrt{2\pi}}
  \ope^{-\frac{(y-X)^2}{2}}
y \dd y}
  \\
  & = & \E[f_X^*]{X \left(1 - \Qf{X}
    - \Qf{\amp-X} \right)}
  \nonumber\\
  && + \E[f_X^*]{ \left(\frac{1}{\sqrt{2\pi}} e^{-\frac{X^2}{2}}  - \frac{1}{\sqrt{2\pi}} e^{-\frac{(\amp-X)^2}{2} }
   \right)}
  \\
  & <   & \E[f_X^*]{X} +  \frac{1}{\sqrt{2\pi}} \cdot \left(1 - e^{-\frac{\amp^2}{2} }  \right),
\end{IEEEeqnarray}
where the last inequality holds because the  $\Qf{\cdot}$-function is positive and the exponential function monotonically increasing.
\end{IEEEproof}

We continue with our upper bound. By plugging these lemmas into \eqref{eq:first_step} and choosing
\begin{equation}
\tau = \frac{  \amp \cdot \zeta_0(\lambda_1,\lambda_2)} {  \amp \cdot \zeta_0(\lambda_1,\lambda_2)+ \sqrt{2 \pi e \sigma^2}},
\end{equation}
we obtain:
\begin{IEEEeqnarray}{rCl}
	\lefteqn{ \C(\alpha_1,\alpha_2,\amp)} \quad \nonumber \\
		 & \leq  & \log \frac{ \amp \cdot \zeta_0(\lambda_1,\lambda_2) }{\tau} \cdot   P_{f_Y^*}(\mathcal{I}) + \log \frac{\sqrt{2 \pi e \sigma^2}}{1-\tau}  \cdot  P_{f_Y^*}(\mathcal{I}^c)\nonumber \\
	 & & + \frac{ \lambda_1 }{\amp}\E[f_X^*]{X}+ \frac{\lambda_2 }{\amp^2}\big(\E[f_X^*]{X^2} +\sigma^2\big)  \nonumber \\
& &+  \frac{\lambda_1}{\amp} \left(1 - \frac{1}{\sqrt{2\pi}} e^{-\frac{\amp^2}{2} } \right) - \frac{1}{2} \log (2 \pi e\sigma^2) \\
& = &  \log \Big(\amp \cdot  \zeta_0(\lambda_1,\lambda_2) +\sqrt{2 \pi e \sigma^2}\Big) \cdot   P_{f_Y^*}(\mathcal{I})  \nonumber \\
& & +  \log \Big(\amp \cdot \zeta_0(\lambda_1,\lambda_2) +\sqrt{2 \pi e \sigma^2}\Big)  \cdot  P_{f_Y^*}(\mathcal{I}^c) \nonumber \\
	 & & + \frac{ \lambda_1 }{\amp}\E[f_X^*]{X}+ \frac{\lambda_2 }{\amp^2}\big(\E[f_X^*]{X^2} +\sigma^2\big)  \nonumber \\
& &+  \frac{\lambda_1}{\amp} \left(1 - \frac{1}{\sqrt{2\pi}} e^{-\frac{\amp^2}{2} } \right) - \frac{1}{2} \log (2 \pi e\sigma^2) \\
& \leq &  \log \Big( \amp \cdot \zeta_0(\lambda_1,\lambda_2) +\sqrt{2 \pi e \sigma^2}\Big)  + \lambda_1 \alpha_1 +\lambda_2 \alpha_2  + \lambda_2 \frac{\sigma^2}{\amp^2}\nonumber \\
 & &+  \frac{\lambda_1}{\amp} \left(1 - \frac{1}{\sqrt{2\pi}} e^{-\frac{\amp^2}{2} } \right) - \frac{1}{2} \log (2 \pi e\sigma^2) \\
 & = &  \log \left( 1+ \frac{\amp \cdot \zeta_0(\lambda_1,\lambda_2) }{2 \pi e \sigma^2}\right)  + \lambda_1 \alpha_1 +\lambda_2 \alpha_2  \nonumber \\
 & &+ \lambda_2 \frac{\sigma^2}{\amp^2}+  \frac{\lambda_1}{\amp} \left(1 - \frac{1}{\sqrt{2\pi}} e^{-\frac{\amp^2}{2} } \right) .
\end{IEEEeqnarray}

We can conclude that for any choice of $\lambda_1,\lambda_2\geq 0$:
\begin{IEEEeqnarray}{rCl}
\lefteqn{\varlimsup_{\amp \to \infty}\left( \C(\alpha_1,\alpha_2,\amp) -\log \frac{\amp }{\sqrt{2 \pi e\sigma^2}}\right)}  \quad  \nonumber \\
 & \leq  & \log  \zeta_0(\lambda_1,\lambda_2)  + \lambda_1 \alpha_1 + \lambda_2 \alpha_2. \label{eq:u}
\end{IEEEeqnarray}

\subsection{Distinction of the Four Cases}
We now show that the case distinction proposed in the theorem partitions the set of all $(\alpha_1, \alpha_2)$-parameters and that the described choice of $\lambda_1^*, \lambda_2^*$-parameters exists in each subset.
More specifically, we  show that  the proposed case distinction coincides with the case distinction  that arises when minimizing the 
right-hand side of \eqref{eq:u}, i.e., the function
\begin{equation}
\Gamma(\lambda_1,\lambda_2): =  \log  \zeta_0(\lambda_1,\lambda_2) + \lambda_1 \alpha_1 + \lambda_2 \alpha_2,
\end{equation}
over the choices $\lambda_1,\lambda_2 >0$,  and we show that the  $\lambda_1^*, \lambda_2^*$ values given in the theorem are  the minimizers of this function.

Consider the partial derivatives of this function:
\begin{subequations}
	\label{eq:partial}
\begin{IEEEeqnarray}{rCl}\label{eq:partial1}
\frac{ \partial \Gamma}{\partial \lambda_1 }= -\frac{ \zeta_1(\lambda_1,\lambda_2)}{ \zeta_0(\lambda_1,\lambda_2)} +\alpha_1
\end{IEEEeqnarray}
and
\begin{IEEEeqnarray}{rCl}\label{eq:partial2}
\frac{ \partial \Gamma}{\partial \lambda_2 }=- \frac{  \zeta_2(\lambda_1,\lambda_2)  }{ \zeta_0(\lambda_1,\lambda_2)} +\alpha_2,
\end{IEEEeqnarray}
\end{subequations} as well as its Hessian matrix
\begin{IEEEeqnarray}{rCl}
\lefteqn{\mathbbm{H} \Gamma (\lambda_1,\lambda_2) :=
\begin{pmatrix}
\frac{\partial^2 \Gamma(\lambda_1,\lambda_2)}{ \partial \lambda_1^2}  &  \frac{\partial^2 \Gamma(\lambda_1,\lambda_2)}{ \partial \lambda_1 \partial \lambda_2 }   \\[2ex]
\frac{\partial^2 \Gamma(\lambda_1,\lambda_2)}{ \partial \lambda_1 \partial \lambda_2 }   & \frac{\partial^2 \Gamma(\lambda_1,\lambda_2)}{ \partial \lambda_2^2}  \end{pmatrix}}  \\
&=& \begin{pmatrix} \zeta_2(\lambda_1,\lambda_2)-  \zeta_1^2(\lambda_1,\lambda_2) & c(\lambda_1,\lambda_2) \\ c(\lambda_1,\lambda_2)& \zeta_4(\lambda_1,\lambda_2)-  \zeta_2^2(\lambda_1,\lambda_2)  \end{pmatrix} ,\IEEEeqnarraynumspace
\end{IEEEeqnarray}
where
\begin{equation}
 c(\lambda_1,\lambda_2):= \zeta_3(\lambda_1,\lambda_2) -  \zeta_1(\lambda_1,\lambda_2)\cdot \zeta_2(\lambda_1,\lambda_2).
\end{equation}
Since for any pair $(\lambda_1,\lambda_2)$ the Hessian  $\mathbbm{H} \Gamma (\lambda_1,\lambda_2)$ is a two-by-two matrix with positive trace and  determinant, all its  eigenvalues are positive, and the Hessian itself  is  positive definite  for all $(\lambda_1,\lambda_2)$. As a consequence, the function $\Gamma(\lambda_1,\lambda_2)$ is  jointly strictly convex in both arguments and the minimizer  $(\lambda_1^*,\lambda_2^*)$ of  $\Gamma(\lambda_1,\lambda_2)$, for $\lambda_1,\lambda_2 \geq 0$ is accordingly obtained  as follows, depending on the values of $\alpha_1$ and $\alpha_2$:
\begin{enumerate}
\item If  both partial derivatives of $\Gamma$ at the origin are strictly positive, i.e.,
\begin{IEEEeqnarray}{rCl}\label{eq:partial0}
-  \frac{ \zeta_1(0,0) }{ \zeta_0(0,0)} +\alpha_1= -\frac{1}{2} + \alpha_1 &>& 0,\\[1.2ex]
  - \frac{\zeta_2(0,0) }{ \zeta_0(0,0)} +\alpha_2 = -\frac{1}{3} + \alpha_2& > & 0,
\end{IEEEeqnarray}
then $\lambda_1^*=\lambda_2^*=0$.
\item If for some $\lambda_1'>0$  the partial derivatives of $\Gamma$ satisfy 
\begin{IEEEeqnarray}{rCl}\label{eq:partial1}
-  \frac{ \zeta_1(\lambda_1',0) }{ \zeta_0(\lambda_1',0)} +\alpha_1  &=& 0,\\[1.2ex]
-  \frac{ \zeta_2(\lambda_1',0) }{ \zeta_0(\lambda_1',0)} + \alpha_2 & > & 0,
\end{IEEEeqnarray}
then $\lambda_1^*=\lambda_1'$ and $\lambda_2^*=0$.
\item  If for some $\lambda_2'>0$  the partial derivatives of $\Gamma$ satisfy 
\begin{IEEEeqnarray}{rCl}\label{eq:partial1}
-  \frac{ \zeta_1(0,\lambda_2') }{ \zeta_0(0,\lambda_2')} +\alpha_1   & >& 0\\[1.2ex]
-  \frac{ \zeta_2(0,\lambda_2') }{ \zeta_0(0,\lambda_2')} + \alpha_2 & =& 0,
\end{IEEEeqnarray}
then $\lambda_1^*=0$ and $\lambda_2^*=\lambda_2'$.

  \item If for some $\lambda_1', \lambda_2'>0$  the partial derivatives of $\Gamma$  at  $(\lambda_1',\lambda_2')$ are both zero, i.e.,
\begin{IEEEeqnarray}{rCl}\label{eq:partial1}
-  \frac{ \zeta_1(\lambda_1',\lambda_2') }{ \zeta_0(\lambda_1',\lambda_2')} +\alpha_1   & =& 0\\[1.3ex]
-  \frac{ \zeta_2(\lambda_1',\lambda_2') }{ \zeta_0(\lambda_1',\lambda_2')} + \alpha_2 & =& 0,
\end{IEEEeqnarray}
then $\lambda_1^*=\lambda_1'$ and $\lambda_2^*=\lambda_2'$. \end{enumerate}
Since the strictly convex function  $\Gamma(\lambda_1, \lambda_2)$ has exactly one minimizing pair, combined with continuity considerations, this concludes the proof of the theorem.

\section{Acknowledgement }
The authors thank Lina Mroueh for interesting discussions.


\end{document}